\documentstyle[aps,prd]{revtex}
\newcommand{\be}{\begin{equation}}
\newcommand{\ee}{\end{equation}}
\newcommand{\ba}{\begin{eqnarray}}
\newcommand{\ea}{\end{eqnarray}}

\def\edth{\raise1.3ex\hbox{$\scriptscriptstyle/$}\mkern-9mu\partial}
\begin{document}

\title{String In Noncommutative Backgrund; Another Approach}
\author{R. Bhattacharyya\footnote{E-address: rajsekhar@vsnl.net}} 
\address{ Department of Physics, Dinabandhu Andrews College, Calcutta 700084,
India}

\maketitle
\vspace{0.5cm}

\begin{abstract}

We propose a model action in 1+1 flat space-time (compact in spatial
dimension) embedded in $D$ flat space-time with a non dynamical space-time dependent two vector. For
the above
constrained system Dirac brackets of suitably defined co-ordinates turn out
to be non zero and under specific choice of the two vector the model action
reduces to a general action of open string in a noncommutative backgrund where we
find the natural embedding of the open string action of [6], more generally with variable
magnetic field, together with the general form of the action which is assumed in
space-time dependent Lagrangian formalism [9]. Further the above analysis
reveals the significant contribution of constant weight factor for Lagrangian
density in space-time noncommutativity $\theta^{\mu\nu}$ although it is 
insignificant for the equations of motion.


{PACS number(s): 11.25.-w~~02.40.Gh}
\end{abstract}

\newpage  

The idea of noncommutative space-time is quite old [1,2]. The quantum phase
space is the first example of such structure for the co-ordinates. In the
last two decades it has been
revived in the study of string theories [3] and in field theories [4] specially
with the emergence of noncommutative Yang-Mills theory in the context of
M-theory compactification in various limits [5] or as low energy limit of
open string ending on a D-brane with constant Nevue-Scwarz $B$ field (which
is also interpreted as the magnetic field on D-brane) [6]. 

Noncommutative geometry has also been analysed in some other physical
theories resulting from the canonical quantisation of constrained systems 
with second class constraints [7], specifically the models of non commutative
relativistic and non-relativistic particles [8] where a Chern-Simons term has 
been added in a first order action. 

Recently in the formalism of space-time dependent Lagrangian [9] (where the
explicit space-time dependence of Lagrangian density is assumed) the
noncommutative structure of co-ordinates arises naturally on the spatial boundary of the
theory both in the analysis of weak-strong duality of 1+1 Sine-Gordon and 
massive Thirring models [9b] and in
context of electro-magnetic duality in $SO(3)$ Yang-Mills theory with
 D-brane structure [9c].  

In the spirit of space-time dependent Lagrangian formalism here we present a 
general model action on 1+1 flat space-time,
compact in its spatial dimension and embedded in $D$ dimensional space-time
with a non dynamical space-time dependent two vector. Analysing this
constrained system we find non zero Dirac brackets of the co-ordinate
variables and for a particular choice of the two vector the above model reduces
to a general open string action in a noncommutative background which not
only embeds
the open string action in [6] but with a variable magnetic field on D-brane.
Further this also recovers the desired assumed form of action in space-time
dependent formalism and also couples the weight factor of Lagrangian
density (which in usual theories is considered as 1 and does not contribute to
the equations of motion [9]) with co-ordinate noncommutativity
${\theta}^{\mu\nu}$, where for noncommutative co-ordinates $X^\mu$ and
$X^\nu$, $$[X^\mu,X^\nu]=i{\theta}^{\mu\nu}$$      

So specifically we propose a model action on a 1+1 flat Minkowski space-time 
(compact in spatial dimension $\sigma$) embeded in $D$ space-time with the
fields ${X_a}^{\mu}(\sigma,\tau)$,
${V_a}^{\mu}(\sigma,\tau)$ and a three indexed antisymmetric tensor 
${\theta}^{a\mu\nu}(\sigma,\tau)$ where
${\theta}^2={\theta}^{a\mu\nu}{\theta}_{a\mu\nu}$. We assume
greek indices $\mu$, $\nu$ for $D$ dimensions where roman indices $a$, $b$ for
the 1+1 embeded dimensions.
$$S_s=-\frac{1}{4\pi{\alpha}'}\int d\tau d\sigma {V_a}^{\mu}{V^a}_{\mu}
-\frac{1}{2\pi{\alpha}'}\int d\tau d\sigma {\Lambda}_b {\partial}^b
{X_a}^{\mu}{V^a}_{\mu} -\frac{1}{2}\int d\tau d\sigma
{\partial}_a {V}_{b\mu}\frac{{\theta}^{a\mu\nu}}{\theta^2}{V^b}_{\nu}\eqno(1)$$ 
Here ${\alpha}'$ is a constant which will later be identified with the
inverse 
string tension and ${\Lambda}_b$ is a non-dynamical space-time dependent two 
vector. It is to be noted that the above action is not of the same form of that in space-time
dependent formalism [9] and unlike them it is also Poncare invariant.

Starting from (1) to go over the Hamiltonian formalism the primary consraints
for the above model are
$${A^a}_{\mu}={P^a}_{\mu}+\frac{{\Lambda}_0
{V^a}_{\mu}}{2\pi{\alpha}'}=0~~~~~;~~~~~
{B^a}_{\mu}={{\Pi}^a}_{\mu}+\frac{{{\theta}^0}_{\mu\nu}{V}^{b\nu}}
{2{\theta}^2}=0~~~~~;~~~~~P_{(\theta)a\mu\nu}=0$$
where ${P^a}_{\mu}$, ${{\Pi}^a}_{\mu}$ and $P_{(\theta)a\mu,\nu}$ are the
conjugate momenta for ${X^a}_{\mu}$, ${V^a}_{\mu}$ and
${\theta}^{a\mu\nu}$. Assuming usual equal time Poisson bracket  
$$\{{X_a}^{\mu}(\sigma,\tau),{P_b}^{\nu}({\sigma}',\tau)\}={\delta}_{ab}
{\delta}^{\mu\nu}\delta(\sigma-{\sigma}')~~~~~;~~~~~
\{{V_a}^{\mu}(\sigma,\tau),{\Pi_b}^{\nu}({\sigma}',\tau)\}={\delta}_{ab}
{\delta}^{\mu\nu}\delta(\sigma-{\sigma}')$$
(where all others are zero and here we do not require the Poisson brackets of
$\theta$'s and
$P_{(\theta)}$'s), we find 
$$\{{A_a}^{\mu}(\sigma,\tau),{A_b}^{\nu}({\sigma}',\tau)\}=0~~~~~;~~~~~
\{{B_a}^{\mu}(\sigma,\tau),{B_b}^{\nu}({\sigma}',\tau)\}=
-\frac{\theta^{0\mu\nu}}{{\theta}^2}{\delta}_{ab} \delta(\sigma-{\sigma}')$$ 
$$\{{A_a}^{\mu}(\sigma,\tau),{B_b}^{\nu}({\sigma}',\tau)\}=
\frac{{\Lambda}_0}{2\pi{\alpha}'}{\delta}_{ab}
{\delta}^{\mu\nu}\delta(\sigma-{\sigma}')$$
and the inverses
$$\{{B_a}^{\mu}(\sigma,\tau),{B_b}^{\nu}({\sigma}',\tau)\}^{-1}=0~~~~~;~~~~~
\{{A_a}^{\mu}(\sigma,\tau),{A_b}^{\nu}({\sigma}',\tau)\}^{-1}=
-\frac{(2\pi{\alpha}')^2\theta^{0\mu\nu}}{{\theta}^2 {\Lambda_0}^2}{\delta}_{ab}
\delta(\sigma-{\sigma}')$$
$$\{{A_a}^{\mu}(\sigma,\tau),{B_b}^{\nu}({\sigma}',\tau)\}^{-1}=
-\frac{2\pi{\alpha}'}{{\Lambda}_0}{\delta}_{ab}
{\delta}^{\mu\nu}\delta(\sigma-{\sigma}')$$
and also the Dirac brackets 
(the genaralisation of Poisson brackets of unconstrained system)  
$$\{{X_a}^{\mu}(\sigma,\tau),{X_b}^{\nu}({\sigma}',\tau)\}_{DB}=
-\frac{(2\pi{\alpha}')^2{\theta^{0\mu\nu}}}{{\theta}^2{\Lambda_0}^2}{\delta}_{ab}
 \delta(\sigma-{\sigma}')$$ 
$$\{{P_a}^{\mu}(\sigma,\tau),{P_b}^{\nu}({\sigma}',\tau)\}_{DB}=0~~~~~;~~~~~
\{{X_a}^{\mu}(\sigma,\tau),{P_b}^{\nu}({\sigma}',\tau)\}_{DB}={\delta}_{ab}
\delta^{\mu\nu}\delta(\sigma-{\sigma}')$$
Defining $\bar{{X_a}^{\mu}}=\frac{1}{l}{\int_0}^l d\sigma {X_a}^{\mu}$ and 
$\bar{{P_a}^{\mu}}={\int_0}^l d\sigma {P_a}^{\mu}$ where $l$ is the length
of the compact dimension we restate the Dirac brackets
$$\{\bar{{X_a}^{\mu}},\bar{{X_b}^{\nu}}\}_{DB}=
-\frac{(2\pi{\alpha}')^2{\theta^{0\mu\nu}}}{l{\theta}^2{\Lambda_0}^2}{\delta}_{ab}$$      
$$\{\bar{{P_a}^{\mu}},\bar{{P_b}^{\nu}}\}_{DB}=0~~~~~;~~~~~
\{\bar{{X_a}^{\mu}},\bar{{P_b}^{\nu}}\}_{DB}={\delta}_{ab}
\delta^{\mu\nu}$$

Now before entering into the Hamiltonian description we break the Poincare
invariance of above action by choosing $\Lambda_1=0$ and
$\Lambda_0=\Lambda$. Here it is to be noted that the above Dirac brackets do
not respect the choice of $\Lambda_1$. So under such choice 
$$H=\frac{1}{4\pi{\alpha}'}\int d\sigma {V_a}^{\mu}{V^a}_{\mu}
+\frac{1}{2}\int d\sigma
{\partial}_1 {V}_{b\mu}\frac{{\theta}^{1\mu\nu}}{\theta^2}{V^b}_{\nu}+{\lambda^{\mu}}_{1a}\int
d\sigma {A^a}_{\mu}+{\lambda^{\mu}}_{2a}\int 
d\sigma {B^a}_{\mu}+\lambda_{(\theta)a\mu\nu}\int d\sigma P^{(\theta)a\mu\nu}\eqno(2)$$ 
where $\lambda_{(\theta)a\mu\nu}$ is undetermined Lagrangian multipliers as 
$P^{(\theta)a\mu\nu}$ is first class constraints and for second class
constraints ${A^a}_{\mu}$, ${B^a}_{\mu}$ we have
$${\lambda^{\mu}}_{1a}=-\frac{{V_a}^{\mu}}{{\Lambda}}
+\frac{\pi{\alpha}'}{{\Lambda}\theta^2}{\theta}^{1\mu\nu}{\partial}_1
{V}_{a\nu}~~~~~;~~~~~{\lambda^{\mu}}_{2a}=0$$
The equations of motion which follows from (2) are
$$\partial_0 {X_a}^{\mu}=-\frac{{V_a}^{\mu}}{{\Lambda}}
+\frac{\pi{\alpha}'}{{\Lambda}\theta^2}{\theta}^{1\mu\nu}{\partial}_1
{V}_{a\nu}~~~~~;~~~~~\partial_0 {P_a}^{\mu}=0\eqno(3)$$
and for two vectors $\vec X^{\mu}$ if we use
$\epsilon^{ab}\partial_a{X_b}^{\mu}=c^{\mu}(\sigma,\tau)$ and
$\partial_a{X_a}^{\mu}=d^{\mu}(\sigma,\tau)$ the equations of motion reduces to
$${V_0}^{\mu}=-(\partial_0 {X_0}^{\mu}){\Lambda}
+\frac{\pi{\alpha}'}{\theta^2}{\theta}^{1\mu\nu}{\partial}_1
{V}_{0\nu}$$
$${V_1}^{\mu}=-(\partial_1 {X_0}^{\mu}){\Lambda}                            
+\frac{\pi{\alpha}'}{\theta^2}{\theta}^{1\mu\nu}{\partial}_1
{V}_{1\nu}+c^{\mu}{\Lambda}$$
Further identifying ${X_0}^{\mu}=x^{\mu}$ the above action (1) reduces
to
$$S=\frac{1}{4\pi{\alpha}'}\int d\tau d\sigma{\Lambda}^2
({\partial_a}x^{\mu}{\partial^a}x_{\mu}
-\frac{1}{2}{\partial}_a x_{\mu}\partial^a
\frac{{\theta}^{b\mu\nu}}{\theta^2}\partial_b x_{\nu}) 
+\bar{S}\eqno(4)$$
where $\bar{S}$ involves terms ${\alpha}'$, ${{\alpha}'}^2$, $c^{\mu}$, $d^{\mu}$
and total derivatives.
The Dirac brackets reduces to
$$\{\bar{{x}^{\mu}},\bar{{x}^{\nu}}\}_{DB}=
-\frac{(2\pi{\alpha'})^2{\theta^{0\mu\nu}}}{l{\theta}^2{\Lambda}^2}$$      
$$\{\bar{{p}^{\mu}},\bar{{p}^{\nu}}\}_{DB}=0~~~~~;~~~~~
\{\bar{{x}^{\mu}},\bar{{p}^{\nu}}\}_{DB}=
\delta^{\mu\nu}$$

So if we identify 1+1 Minkowski space (compact in $\sigma$ direction) as the world sheet of the
string in orthonormal gauge with $\frac{1}{2\pi\alpha'}$ as the string tension
then (4) represents the general action for open string in noncommutative
background where we can
identify the embedding of the action of open string in presence of magnetic
field on D$p$-brane. For $\Lambda=1$ 
one finds $S$ is a Poincare invariant action and having no past history. Thus we can argue that $S_s$ is
a more general action for the open string in noncommutative back ground
where after the breaking of the Poincare invariance by above choice of two vector we
get the action in [6] embedded in $S$ which is Poincare invariant. Further $S$ is
also general in that sense as it includes the non constant magnetic field
which arises naturally in the action due to the $\theta$ field, as it couples the
indices of space-time with that of world sheet. We can identify the magnetic
field by
$$B^{\mu\nu}=\frac{1}{2}\epsilon_{ab}\partial^a
\left(\frac{{\theta^{b\mu\nu}}}{\theta^2}\right)$$ 
and under the choice of ${\theta^{0\mu\nu}}$= constant, magnetic field can still
be in the action as long as ${\theta^{1\mu\nu}}$ is a non constant field of time
and thus ${\theta^{1\mu\nu}}$ can produce a smooth change of magnetic field.

Now if we turn to Dirac brackets for the choice
of ${\theta^{0\mu\nu}}$= constant, we get usual noncommutative relations with
the significant contributions of constant $\Lambda$ and  
${\alpha'}^2$. So constant
$\Lambda$ becomes non trivial for the space-time noncommutativity although it does not
contribute to the equations of motion.

Another interesting thing is the revival of the general form of the action
of the space-time dependent Lagrangian formalism. In that formalism for sake
of more general Lagrangian density explicit 
space-time dependence is included in it so also in the
action. Precisely, in $D$ space-time dimensions and for any arbitrary field $f$ it is assumed that Lagrangian
density $L'(f,\partial f,x)=\rho(x)L(f,\partial f)$ where $\rho(x)$ is weight
factor and the action corresponding to $L'$ is $S'=\int d^Dx\rho(x)L$. So in
formalism [9], the open string action in [6] would take the form 
$$S'=\frac{1}{4\pi{\alpha}'}\int d\tau d\sigma\rho({\partial_a}x^{\mu}{\partial^a}x_{\mu}
-\frac{1}{2}{\partial}_a x_{\mu}\epsilon^{ab}B^{\mu\nu}\partial_b x_{\nu})$$ 
but to start with, we do not assume this form for $S_s$ but as it reduces to $S$ we just
recover that form.   

The action $S_s$ needs further investigation for the following reasons,
firstly whether it contains any other theory for a new choice of two vector
and secondly what may be the physical theory (if any such theory exists at
all) whose acton is $S_s$. We can also study $S$ as it includes more
ineraction terms than that of [6].

Lastly for $\Lambda=1$ we find $S$, a Poincare invariant action after
breaking that of $S_s$. We can proceed in the reverse direction i.e starting
from any Poincare invariant usual physical theory analogous to $S$ for
$\Lambda=1$, we can analyse whether we can get hold of another Poincare
invariant theory analogous of $S_s$ and if for several such $S$'s, $S_s$ is unique
then this may be a way for uniting the physical theories.

{\bf  Acknowledgement}\\
I am greatful to D.Gangopadhyay for useful discussion.


\begin{thebibliography}{1000}

\bibitem{} P.A.M. Dirac, Proc. Roy. Soc. A109(1926)642; Proc. Camb. Phil.
Soc. 23(1926)412 
\bibitem{} H.S. Snyder, Phys. Rev. 71(1947)38; Phys. Rev. 72(1947)68
\bibitem{} E. Witten, Nucl. Phys. B 268 253(1986)
\bibitem{} A. Connes, Non-commutative Geometry, Academic Press(1994)
\bibitem{} A. Connes, M.R. Douglas and A. Schwarz, JHEP 9802 003(1998)
\bibitem{} N. Seiberg and E. Witten, JHEP 9909 032(1999)
\bibitem{} M. Henneaux and C. Tietelboim, Quantisation of Gauge System,
Princeton University Press
\bibitem{} A.A. Deriglazov, Phys. Lett. B 530(2002)235; hep-th/0207274
\bibitem{} R. Bhattacharyya and D. Gangopadhyay, Mod. Phys. Lett.
A15,901(2000); Mod. Phys. Lett.
A17,729(2002); hep-th/0210051; D. Gangopadhyay, R. Bhattacharyya and 
L. P. Singh, hep-th/0208097



\end{thebibliography}
\end{document}